\documentclass{sig-alternate-nocopyright}
\usepackage{booktabs}
\usepackage{url}
\usepackage{latexsym}

\usepackage{epsfig}
\usepackage{subfigure}
\usepackage{xcolor}
\usepackage{amsmath,amssymb,color}
\usepackage{amssymb,color}
\usepackage{array}
\usepackage{algorithm}
\usepackage[noend]{algorithmic}
\usepackage{multirow}
\usepackage{graphics,epsfig}

\newcommand{\cut}[1]{}
\newcommand{\TODO}[1]{}

\begin{document}
\title{LinkedIn Salary: A System for Secure Collection and Presentation of Structured Compensation Insights to Job Seekers}

\numberofauthors{1}
\author{\alignauthor 
Krishnaram Kenthapadi \ \
Ahsan Chudhary \ \
Stuart Ambler\\
\affaddr{LinkedIn Corp}\\
\affaddr{Sunnyvale, CA, USA}\\
       \email{\{kkenthapadi, achudhary, sambler\}@linkedin.com}
}
\maketitle

\begin{abstract}
Online professional social networks such as LinkedIn have enhanced the ability of job seekers to discover and assess career opportunities, and the ability of job providers to discover and assess potential candidates. For most job seekers, salary (or broadly compensation) is a crucial consideration in choosing a new job. At the same time, job seekers face challenges in learning the compensation associated with different jobs, given the sensitive nature of compensation data and the dearth of reliable sources containing compensation data. Towards the goal of helping the world's professionals optimize their earning potential through salary transparency, we present LinkedIn Salary, a system for collecting compensation information from LinkedIn members and providing compensation insights to job seekers. We present the overall design and architecture, and describe the key components needed for the secure collection, de-identification, and processing of compensation data, focusing on the unique challenges associated with privacy and security. We perform an experimental study with more than one year of compensation submission history data collected from over 1.5 million LinkedIn members, thereby demonstrating the tradeoffs between privacy and modeling needs. We also highlight the lessons learned from the production deployment of this system at LinkedIn.
\end{abstract}

\section{Introduction}\label{sec:intro}
Online professional social networks such as LinkedIn have enhanced the ability of job seekers to discover and assess career opportunities, and the ability of job providers to discover and assess potential candidates. 
Compensation is known to be a crucial consideration in choosing a new job opportunity for most job seekers.\footnote{More candidates (74\%) would like to see compensation compared to any other feature in a job posting, according to a survey of over 5000 job seekers in US and Canada~\cite{careerBuilderSurvey2016}. Compensation is valued the most by job seekers when looking for new opportunities, according to a US survey of 2305 adults~\cite{jobSeekerNationStudy2016}.} However, job seekers encounter obstacles in learning the compensation associated with different jobs, given the sensitive nature of compensation data and the dearth of reliable sources containing compensation data. The recently launched LinkedIn Salary product\footnote{\url{https://www.linkedin.com/salary}} has been designed to help job seekers explore compensation along different dimensions, make more informed career decisions, and thereby optimize their earning potential through salary transparency.

With over 500 million members (registered users), together with the associated structured information including the work experience, educational history, and skills, LinkedIn is in a unique position to collect compensation data from its members at scale and provide rich, robust insights covering different aspects of compensation, while preserving member privacy. For instance, we can provide insights on the distribution of base salary, bonus, equity, and other types of compensation for a given profession; how they vary based on factors such as region, experience, education, company size, and industry; and which regions, industries, or companies pay the most.

Besides helping job seekers understand their economic value in the marketplace, the compensation data has several other social benefits. It can help us better understand the monetary dimensions of the Economic Graph~\cite{Wei12} (which includes companies, industries, regions, jobs, skills, educational institutions, etc.). The availability of compensation insights with respect to dimensions such as gender, ethnicity, and other demographic factors can lead to greater transparency, shedding light on the extent of compensation disparity, and thereby help stakeholders including employers, employees, and policy makers to take steps to address pay inequality. Further, products such as LinkedIn Salary can lead to greater efficiency in the labor marketplace by reducing asymmetry of compensation knowledge, and by serving as market-perfecting tools for workers and employers~\cite{Har16}. Finally, such tools have the potential to help students make good career choices, taking expected compensation into account, and to encourage workers to learn skills that are necessary for obtaining well paying jobs, thereby helping narrow the skills gap.

In this paper, we present LinkedIn Salary, a system for securely collecting compensation information from LinkedIn members and providing structured compensation insights to job seekers. We highlight the unique challenges associated with privacy and security while designing and implementing the system, and present the overall design and architecture that incorporates a combination of techniques such as encryption, access control, de-identification, aggregation, and thresholding. We also describe the key components needed for the secure collection, de-identification, and processing of compensation data. We empirically investigate the tradeoffs between privacy and modeling needs using more than one year of compensation submission history data collected from over 1.5 million LinkedIn members. We also highlight the lessons learned in practice from the deployment of this system at LinkedIn.

\section{Problem Setting}\label{sec:problem}
\begin{figure*}
\centering
\includegraphics[width=0.7\textwidth]{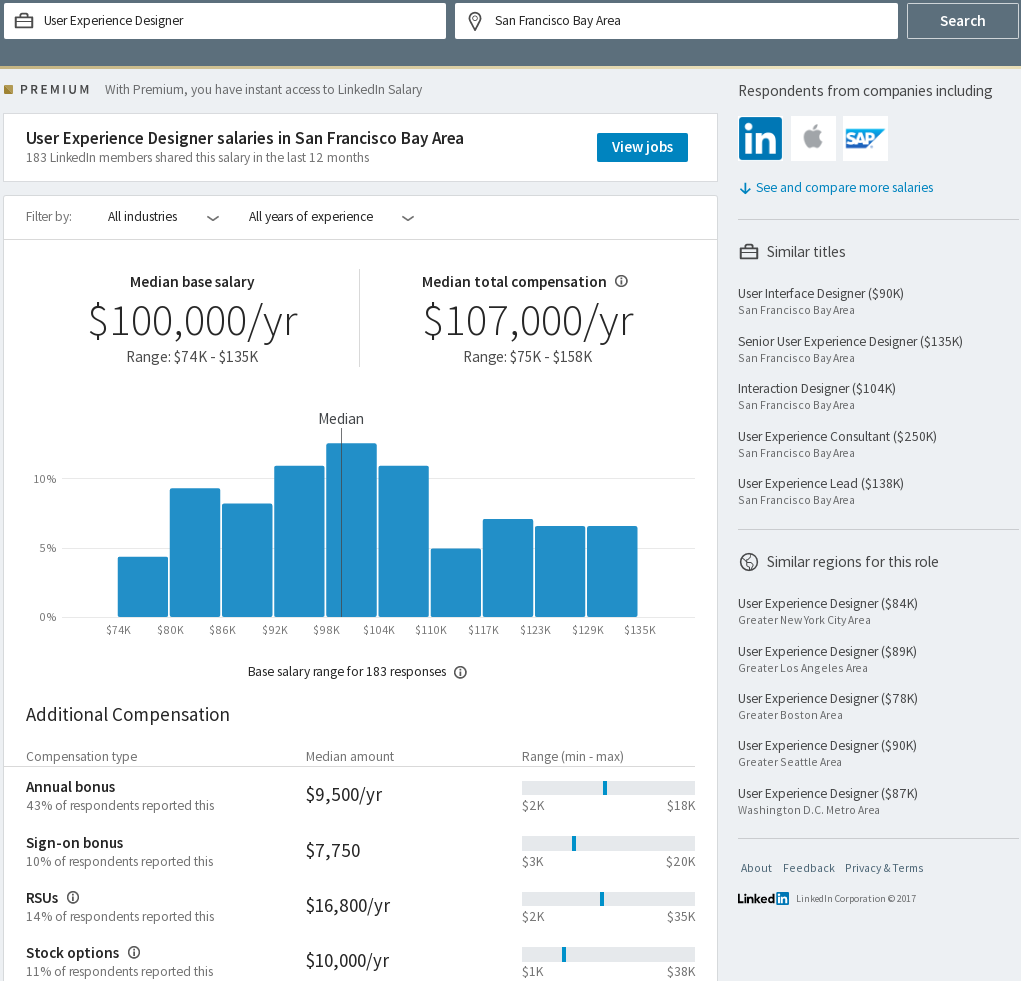}
\caption{LinkedIn Salary Insights Page}
\label{fig:salaryInsight}
\end{figure*}

We next give a brief overview of the product, followed by a discussion of the key system design goals, as well the challenges from the privacy, security, and modeling perspectives. We then present the problem statement.

In the publicly launched LinkedIn Salary product~\cite{salaryBlogPost2016}, users can search for different titles and regions, and explore the corresponding compensation insights (Figure~\ref{fig:salaryInsight}). For a given title and region, we present the quantiles ($10$th and $90$th percentiles, median) and histograms for base salary, bonus, and other types of compensation. We also present more granular insights on how the pay varies based on factors such as region, experience, education, company size, and industry, and which regions, industries, or companies pay the most.

\begin{figure*}
\centering
\includegraphics[width=0.8\textwidth]{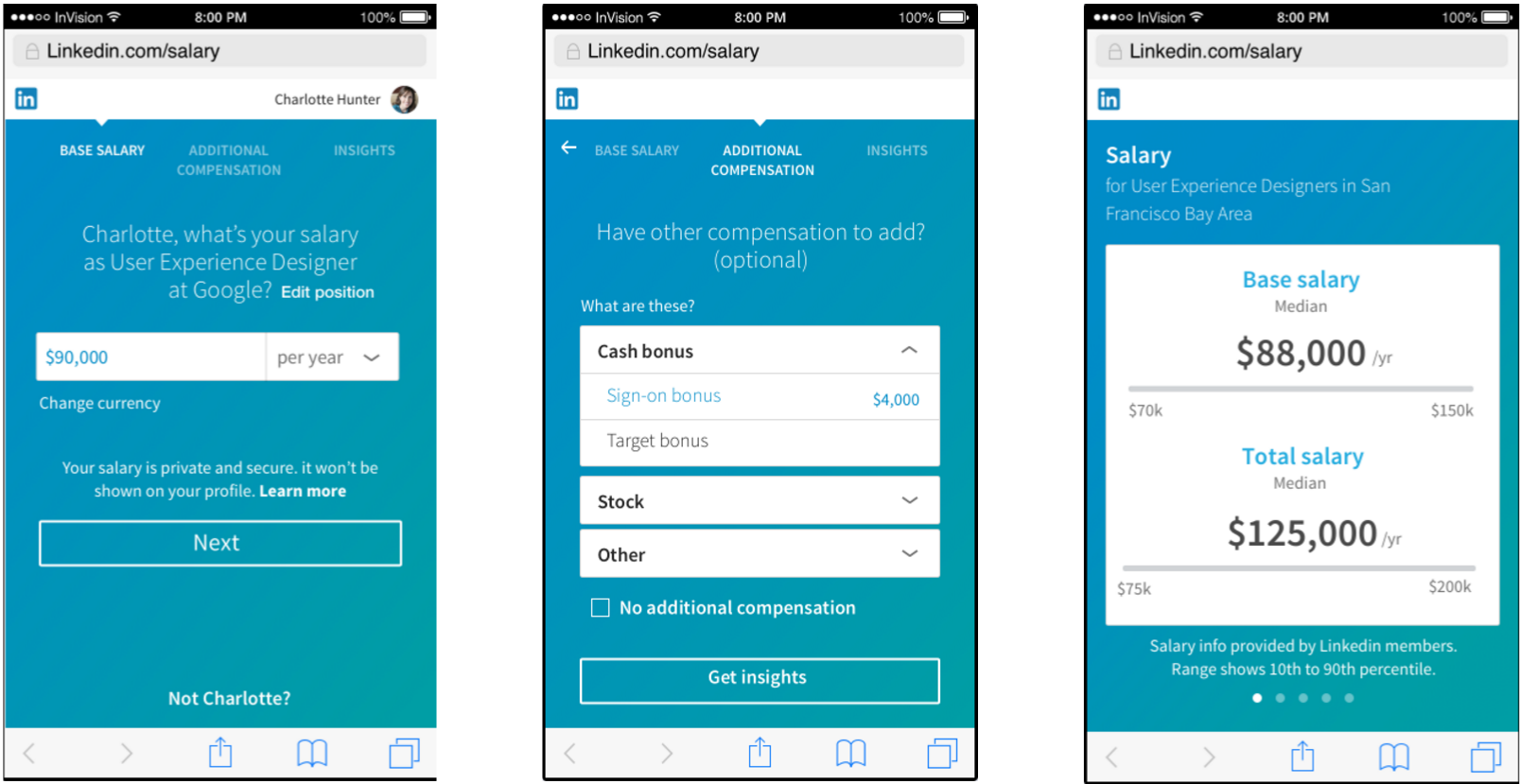}
\caption{LinkedIn Salary Collection Interface. The interfaces for a member to enter the base salary and additional compensation for his/her job position are shown in the first two screens. The compensation insights for the corresponding cohort are shown in the third screen.}
\label{fig:salaryCollection}
\end{figure*}

The compensation insights shown in the product are based on more than one year of compensation data that we have been collecting from LinkedIn members (Figure~\ref{fig:salaryCollection}). We designed the data collection process based on a give-to-get model. First, we select cohorts (such as User Experience Designers in San Francisco Bay Area) with a sufficient number of LinkedIn members. Within each cohort, we send emails to a random subset of members, requesting them to submit their compensation data (in return for aggregated compensation insights later). Once we collect enough data, we get back to the responding members with the compensation insights, and also reach out to the remaining members in those cohorts, promising corresponding insights immediately upon submission of their compensation data. We describe the collection mechanism in further detail in \S\ref{sec:targeting}.

Figure~\ref{fig:salaryCollection} shows screenshots of the collection interface as seen by a hypothetical user. The first two screens show the interface for a member to enter the base salary and additional compensation respectively for his/her current job position. The current job position information is automatically retrieved from the member's profile and displayed on the first screen, with an option for the member to modify if desired. The member can choose to enter different types of additional compensation, as shown in the second screen. After the member submits the compensation data and once the corresponding insights are available, the member is presented with these insights, along with other useful insights such as the compensation for related positions in the same region and the compensation for the same position in related regions.

\subsection{System Requirements and Challenges}\label{sec:challenges}
{\em Preserving privacy of members}: Considering the sensitive nature of compensation data and the desire for preserving privacy of users, a key requirement is to design our system such that there is protection against data breach, and any one individual's compensation data cannot be inferred by observing the outputs of the system. We would like protection against external data breach as well as insider attacks. Further, we require the compensation insights to be generated based on only cohort level data containing de-identified compensation submissions (e.g., salaries for UX Designers in San Francisco Bay Area), limited to those cohorts having at least a minimum number of entries.

{\em Engineering requirements}: The system needs to be designed such that ``bootstrapping'' can be performed when needed. Bootstrapping refers to the ability to regenerate the historical output data whenever there are changes in parts of the system. For example, the need for this functionality could arise due to software bugs in the system implementation, changes to the various services that are invoked by the system, and changes to the product design such as addition of new types of insights.

{\em Modeling on aggregated data}: Due to the privacy requirements stated above, the salary modeling system has access only to cohort level data containing aggregated compensation submissions.
Each {\em cohort} is defined by a combination of attributes such as title, country, region, company, years of experience, and so forth, and contains aggregated compensation entries obtained from individuals having the same values of those attributes.
Within a cohort, each individual entry consists of values for different compensation types such as base salary, annual bonus, sign-on bonus, commission, annual monetary value of vested stocks, and tips, and is available without associated user name, id, or any attributes other than those that define the cohort. As a result, our modeling choices are limited since we have access only to the aggregated data, and cannot, for instance, train prediction models that make use of more discriminating features not available due to de-identification.

{\em Evaluation}: We face unique evaluation and data quality challenges that are typically not present in several other user-facing products such as movie and job recommendations. Since users themselves may not have a good perception of the true compensation range, we cannot perform online A/B testing to compare the compensation insights generated by different models. Further, there are very few reliable and easily available ground truth datasets in the compensation domain, and even when available (e.g., BLS OES dataset~\cite{BlsOes08}), mapping such datasets to LinkedIn's taxonomy is inevitably noisy.

{\em Outlier detection}: The quality of the insights depends on the quality of submitted data, and hence we need to detect and prune potential outlier entries. Such entries could arise due to either mistakes/misunderstandings during submission, or intentional falsification (such as someone attempting to game the system). We required a solution that would work even during the early stages of data collection, when this problem was more challenging, and there may not be sufficient data across say, related cohorts.

{\em Robustness and stability}: Although some cohorts may each have a large sample size, a large number of cohorts typically contain very few (e.g., $< 20$) data points each. Given the desire to have insights for as many cohorts as possible, we need to make sure that the compensation insights are robust and stable even when there is data sparsity. In other words, the insights should be reliable, and not too sensitive to the addition of a new entry for such cohorts. A related challenge is whether we can reliably infer the insights for cohorts with no data at all.

\subsection{Problem Statement}\label{sec:statement}
Our focus in this paper is on the unique challenges associated with privacy and security while designing and implementing the system. Our methodology for addressing the modeling challenges is described in~\cite{KAZA17}. Our problem can thus be stated as follows: {\em How do we design LinkedIn Salary system to meet the immediate and future needs of LinkedIn Salary and other LinkedIn products? How do we design our system taking into account the unique privacy and security challenges, while addressing the product requirements?} We address these questions in \S\ref{sec:arch}, wherein we present the system architecture that incorporates a combination of techniques such as encryption, access control, de-identification, aggregation, and thresholding.
 \section{LinkedIn Salary System Design and Architecture}\label{sec:arch}
\begin{figure*}
\centering
\includegraphics[width=0.7\textwidth]{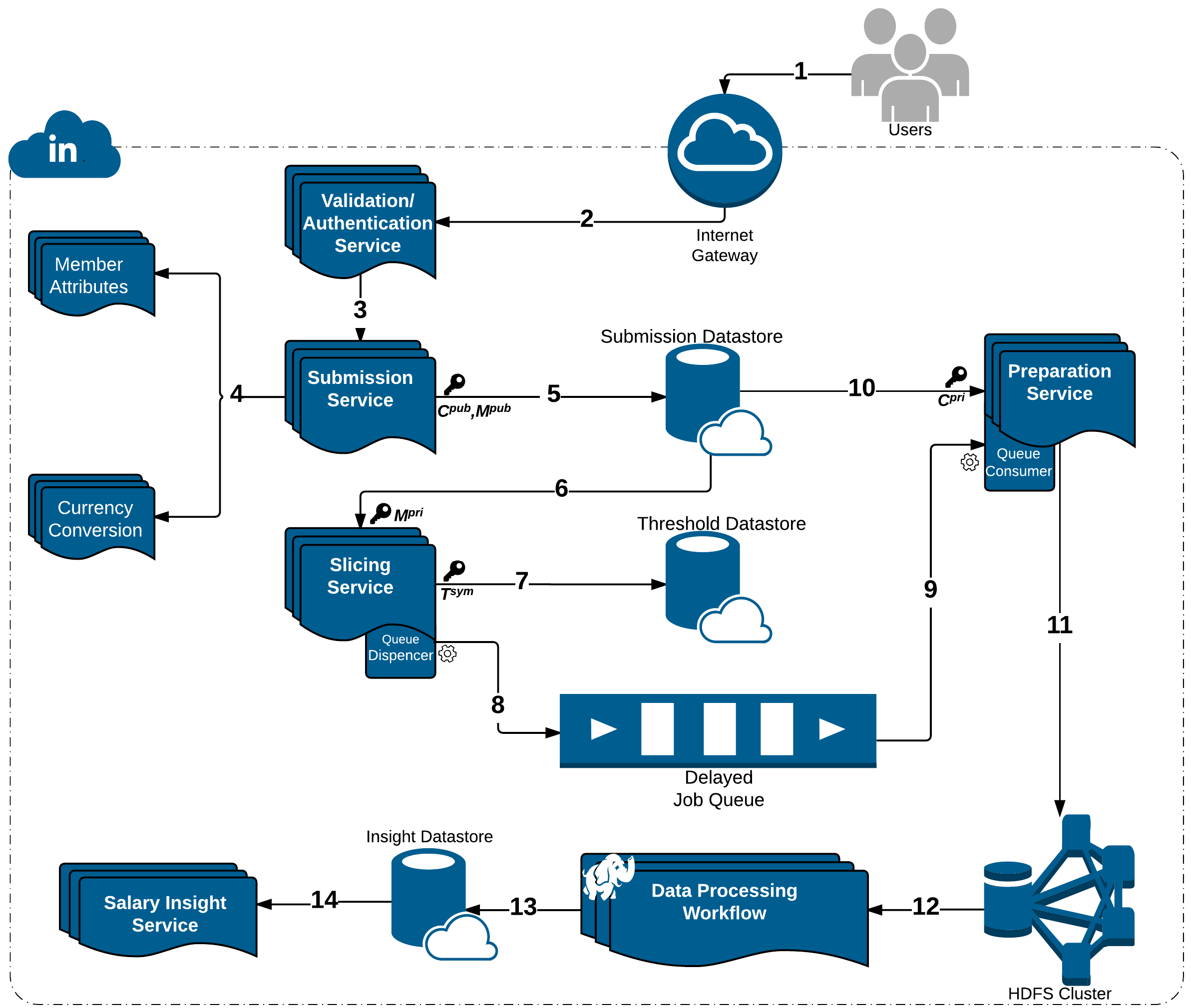}
\caption{LinkedIn Salary System Architecture, consisting of components/services pertaining to collection \& storage, de-identification \& grouping, and insights \& modeling.}
\label{fig:braavosarch}
\end{figure*}

We describe the overall design and architecture of the deployed system that powers the LinkedIn Salary product. As part of the description, we present the security and de-identification mechanisms.
Our system uses a service oriented architecture (see Figure~\ref{fig:braavosarch}), and consists of the following three key components: a collection and storage component, a de-identification and grouping component, and an insights and modeling component. We provide an overview of the requests and responses of the underlying services associated with a member submitting the compensation data to see the insights. We refer to the step numbers found in Figure~\ref{fig:braavosarch} as part of this description.

\subsection{Collection and Storage}\label{sec:collection}
The collection and storage component of our system is responsible for allowing members to submit their compensation information, collecting different member attributes, and securely and privately storing the member attributes as well as the submitted compensation data. The primary entry point for the collection flow is through email campaigns that target LinkedIn members, requesting them to submit their compensation information (described in detail in \S\ref{sec:targeting}). Upon clicking on the email, the member is first authenticated and then taken to the secure web interface (that uses TLS) shown in Figure~\ref{fig:salaryCollection}. Once the compensation data is submitted by the member, this data passes through multiple routing layers before reaching the front-end service. All the intermediate layers also use TLS to secure the data (Step 1 of the architecture diagram). The front-end service is responsible for validation and authentication of the request (Step 2). The data is then handed over to the submission service for storage. The submission service collects member attributes such as title, company, and region, encrypts them along with compensation data and stores in a persistent data store.

{\noindent \em Collection APIs}: LinkedIn exposes all of its services internally using RESTful APIs~\cite{fielding2000architectural}. These services are powered by rest.li, which is a open source REST $+$ JSON framework for building robust, scalable service architectures using type-safe bindings, dynamic discovery, and simple asynchronous APIs~\cite{restli}. In particular, compensation collection APIs also use the rest.li framework. The submission service is exposed as a rest.li endpoint and the front-end service interacts with the submission API over TLS (Step 3 in the diagram). The submission service then extracts attributes from the member profile using relevant data service APIs (Step 4). 
This information is extracted during the time of submission since the member may update the profile over time and we would like to obtain the attribute values as of the time of the compensation submission.
Examples of the extracted member attributes include the standardized (or canonical) versions of job title, company, and region, obtained by invoking respective LinkedIn standardization services.
We also obtain attributes such as the the currency exchange rate at the time of submission.

Once we obtain the member attributes by invoking various LinkedIn Data services, we perform 2048-bit RSA based encryption of compensation data as well as member attributes, and store in our data store using a write-only API (Step 5). We use different sets of public/private keys for encryption/decryption of member attributes and compensation data due to reasons explained in \S\ref{sec:security}.

Our system also provides a verification service which is used to verify if a member has submitted compensation information in the past (not shown in Figure~\ref{fig:braavosarch}). This service is powered by a different data store that only stores whether (and when) a member submitted compensation information, but not the compensation data itself. This store is needed to power the {\em give-to-get} model used by our product.

\subsection{De-identification and Grouping}\label{sec:grouping}
Once the data is stored, we need to prepare it for analysis. We used an approach (inspired by $k$-Anonymity~\cite{samarati1998protecting,samarati2001protecting,Swe02}) which allowed us to use existing LinkedIn infrastructure while preserving the privacy of LinkedIn members. We construct collections of member-submitted compensation information by grouping together members satisfying different predicates, and call these collections as ``cohorts'' or ``slices''. Each {\em cohort} or {\em slice} is defined by a combination of attributes such as canonical title, country, region, company, and years of experience, and contains aggregated compensation entries obtained from individuals having the same values of those attributes. Within a cohort, each individual entry contains values for different compensation types such as base salary, annual bonus, sign-on bonus, commission, annual monetary value of vested stocks, and tips, and is available without associated user name, id, or any attributes other than those that define the cohort. An example of a title-country-region slice would be ``User Experience Designers in San Francisco Bay Area'', and an example of a title-company-country-region slice would be ``User Experience Designers at Google in San Francisco Bay Area''. In addition to the de-identification, we also require that each cohort contain at least a minimum number $k$ of entries, before it is made available for offline data processing and analysis. Note that prior to the above grouping, we apply LinkedIn standardization software to map a free-form attribute to its canonical version. For example, an arbitrary title is mapped to one of about $25,000$ LinkedIn standardized (canonical) titles, and similarly, each company and each region also get mapped to their canonical versions. This mapping is analogous to the generalization step in $k$-Anonymity~\cite{samarati1998protecting,samarati2001protecting,Swe02}.

At first, we wanted to store and process the the encrypted (but sensitive) data containing member attributes and member compensation in LinkedIn's Hadoop ecosystem. However, due to the inherent security limitations of Hadoop and HDFS, we decided to limit the amount of data made available in HDFS towards minimizing the chances of re-identification, and pursued the de-identification approach of slicing and requiring a minimum threshold of entries for processing.

The slicing service runs asynchronously from the submission service, and is triggered for each new submission by a member. Slicing service makes use of Databus, LinkedIn's open-source distributed change-data capture and notification system~\cite{databus} (Step 6 in Figure~\ref{fig:braavosarch}). We limit the data that is stored at the Databus servers to only member attribute information and submission identifiers (but not the compensation data). This ensures that the compensation data is secure even if the Databus service is breached. We use stored attributes to generate the slices, which are then used for analysis and generating insights.

As stated before, we perform analysis on a slice only after it contains a minimum number of submission entries. Our system allows the specification of different minimum thresholds for different slice types (e.g., title-country-region vs title-company-country-region). 
The slicing service uses a distributed data platform to keep track of the thresholds (Step 7). 
The slicing service keeps track of the number of submissions that have been collected so far for each slice, and will decide to make a slice available for processing by subsequent components only once the number of submissions exceeds the minimum threshold. In particular, since the slicing service does not access the compensation data at all, a slice that has not met the threshold will not have the associated compensation entries. Thus, for example, the slice, ``User Experience Designers at Google in San Francisco Bay Area'' would get processed once the threshold is met, while the slice, ``CEOs at LinkedIn in San Francisco Bay Area'' would never get processed and hence the associated compensation will not be available downstream.

Our system supports a mechanism for modifying the timestamp associated with the compensation submission so as to prevent timestamp based inference attacks, that is, prevent the ability to obtain the member's identity by joining with page view or other logs (discussed in detail in \S\ref{sec:timestampjoin}). The modification can be either in the form of a random delay, or based on the other submissions in the cohort. Such modification of the timestamp is performed using the delayed job queue (Step 8).

Once the queue task is ready to be processed, it is picked up by the preparation service (Step 9). As our gateway service to the offline system, the preparation service prepares the data by fetching the compensation data stored during collection (Step 10), and associates it with the slice data. The compensation data corresponding to the submission identifiers in a slice is fetched and decrypted, and afterwards, the submission identifiers are removed from the slice. In this manner, we do not retain any association between the (decrypted) compensation data in a slice and the corresponding submission identifiers. The prepared data is then copied to HDFS for offline processing (Step 11). At this stage, our system supports the ability to perform rounding or generalization of compensation entry values with a goal of minimizing the risk of joining the de-identified data with itself based on the exact compensation value and thereby identifying multiple attributes associated with each compensation entry. 
We remark that the use of two different services, slicing service and preparation service, instead of a combined single service, provides an extra layer of security. This gives us protection against linking together member attribute data and compensation data if any one service gets breached.

\subsection{Compensation Insights and Modeling}
\subsubsection{Selection and Targeting of Members}\label{sec:targeting}
We next describe the data collection process, focusing on the selection of cohorts and members for sending emails requesting submission of compensation data. First, cohorts with a sufficient number of LinkedIn members were selected. The members within a cohort were randomly ordered, and emails were sent to a random subset of members (following this order), requesting them to submit their compensation data (in return for aggregated compensation insights later). Once we collected suffficient data, we got back to the responding members with the compensation insights, and also reached out to the remaining members in the cohort, promising insights immediately upon submission of their compensation data. We partitioned a cohort in this manner since providing compensation insights immediately upon submission results in a higher response rate, as well as a better user experience, and hence we wanted to ensure that as few members as possible were requested prior to the availability of compensation insights.

We dynamically adapted the above process based on the observed response rate. For example, by sending emails to the first $r_1$ members in the random ordering and observing the number of responses $s_1$, we estimated the response rate, $\gamma$ as $s_1/r_1$. If the desired number of responses is $\alpha$, we then reached out to about $r_2 = \alpha/\gamma - r_1$ additional members. Further, we wanted to target a cohort only if we expected enough responses to be able to get back with the compensation insights. For this purpose, we estimated the response rate for a cohort based on other similar cohorts (e.g., considering the response rates for the same title in other regions in the case of a title-country-region cohort).

We remark that through random sampling (which is a special case of probability-based sampling~\cite{groves2011survey}) and targeted data collection, we minimize any selection bias and ensure that the chosen set of members is representative of the underlying population of LinkedIn members in the cohort. In contrast, compensation insights obtained based on self-reported surveys need not be representative of the underlying population. Addressing other types of biases (such as response bias) is an avenue for future work.

\subsubsection{Computation and Presentation of Compensation Insights}\label{sec:modeling}
Whenever compensation insights are requested (say, right after the member entering compensation data, or as part of LinkedIn Salary product), the front-end service queries the verification service via a REST API to determine whether the member is eligible to view the insights. Based on the product and business needs, the eligibility can be defined in terms of criteria such as whether the member has provided his/her compensation data within the last one year (give-to-get model), whether the member has purchased a LinkedIn premium membership, whether the member satisfies a predicate expressed in terms of demographic  / profile-based attributes (e.g., a middle-skilled worker; a student; based in emerging markets), or whether the member satisfies a predicate expressed in terms of activity attributes (e.g., likely to transition from being an active member to an inactive member). If yes, the front-end obtains the compensation insights by querying the salary insight service. For each cohort, depending on data availability, the compensation insight could include the quantiles ($10$th and $90$the percentiles, median), and histograms for base salary and other compensation types. In addition, there are also other insights such as how the pay varies based on factors such as region, industry, experience, education, and company size~\cite{KAZA17}.

The insights are generated using an offline data processing workflow that consumes the de-identified compensation dataset (Step 12 in Figure~\ref{fig:braavosarch}) and populates the insight Voldemort data stores~\cite{sumbaly2012serving} (Step 13), which are then used by the salary insight service (Step 14). Statistical modeling components such as outlier detection and Bayesian hierarchical smoothing are used in the offline workflow to compute robust compensation insights. See~\cite{KAZA17} for an in-depth description of the salary insight service, the offline workflow, and the statistical modeling components.

We remark that in addition to helping compute robust compensation insights for cohorts with very little data, Bayesian smoothing methodology also helps to achieve privacy in certain cases. It is possible for the salaries to be identical in a cohort with very few entries, thereby defeating the benefit of having a minimum threshold after de-identification. For example, the cohort ``User Experience Designers at Google in San Francisco Bay Area'' may contain submissions from just 6 members, each with $110$K as the base salary. In such cases, the smoothing helps to derive a more reliable cohort estimate by ``borrowing strength'' from the ancestral cohorts that have sufficient data, and thereby not reveal the empirical percentiles (all of which equal $110$K)~\cite{KAZA17}. Further, the smoothing helps to minimize the privacy risk by not revealing the empirical percentiles based on very few data points, which could be quite sensitive to the addition of new data points.

\subsection{Security Mechanisms}\label{sec:security}
We next describe the security mechanisms in our architecture.\\

{\noindent \em Encryption (and decryption) of member attributes and compensation data using different sets of public (and private) key pairs}:
Recall that the data collected by the submission service corresponding to each member submission consists of two components: the member attributes and the compensation data. Since the combination of different member attributes could uniquely identify an individual in certain cases~\cite{sweeney2000uniqueness} (e.g., the attributes, title: ``CEO'' and company: ``LinkedIn'' together correspond to one person at a given point of time), we would like to not only store the member attributes and the compensation data in encrypted form, but also use different encryption keys for both. As a result, an attacker that has access to only one of the decryption keys cannot infer the association between the member attributes and the corresponding compensation data. Further, our system is designed in such a way that no service has a need to simultaneously decrypt both the member attributes and the compensation data, and hence is not given access to both decryption keys as elaborated below. By using asymmetric, public key cryptography, we also ensure that the submission service can perform encryption but not decryption. In the very unlikely event of the submission service being breached, the attacker would still not be able to decrypt the historical submissions (since it does not have the private keys), let alone retrieve the historical data (since the access to the store is using a write-only API).\\

{\noindent \em Separation of processing of member attributes and compensation data}:
In our system, the member attributes and the compensation data are never processed at the same time, or by the same service, after the initial submission. This design, together with the use of two different encryption mechanisms, ensures that an attacker would have to break into both encryption systems (and hence multiple services) to be able to connect the dots between member attributes and their corresponding compensation data.\\

{\noindent \em Key store security}:
We achieved key store security by separating the keys into multiple keystores. Although the keystores are password protected, we decided to restrict access to the key resource(s) among different services, and hence separated the keys into multiple keystores.\\

{\noindent \em Limiting access to keys}:
Each service only has access to the public key(s) needed for encryption, and the private key(s) needed for decryption. We next describe the encryption and decryption of data through the system workflow to illustrate this design choice.
Upon obtaining the member attributes and the compensation data, the submission service encrypts them using the respective public keys, $M^{pub}$ and $C^{pub}$, and then stores in the submission data store. Since this service would never be required to decrypt these data, it does not have access to the corresponding private keys. Then, the slicing service which listens to the Databus stream gets triggered. Since this service requires member attribute information to create the slices, it decrypts the member attributes using the corresponding private key, $M^{pri}$. Since the slicing service does not need to access the compensation data at all, it does not possess the corresponding private key, $C^{pri}$. Next, the distributed threshold data store is used to identify whether a slice has met its threshold requirement. The data in this store is encrypted using the symmetric key, $T^{sym}$ instead of using asymmetric, public/private keys since this data is always encrypted and decrypted by the same service (slicing service). Once the threshold for a certain slice has been met, the data is decrypted and pushed into the delayed job queue along with the corresponding slice information. When the preparation service consumes the data from the queue, it fetches the compensation data corresponding to the submission identifiers in each slice from the submission data store and decrypts it using the corresponding private key, $C^{pri}$, prior to copying the data to HDFS. Note that this service does not need to access the member attributes, and hence does not possess the corresponding private key, $M^{pri}$.\\

{\noindent \em Key rotation}:
Our design supports a key rotation mechanism to rotate expired or compromised keys and re-encrypt the data if needed, and to perform periodic re-encryption if needed. This process will be carried out from the service that has access to the private key, e.g., the slicing service would rotate keys for the member attributes, while the preparation service would rotate keys for the compensation data.\\

{\noindent \em No single point of failure}:
By appropriate separation of services and using different sets of encryption/decryption keys for different data components, our design ensures that there is no single point of failure. Member privacy would be preserved if there is any single point of breach to the services or the data stores, for example, if any one system (e.g., HDFS, submission data store, threshold data store, slicing service, preparation service) were to be breached.\\

{\noindent \em Infrastructure security}: In addition to measures for member data privacy, we also focus on the security of the underlying communication and storage infrastructure. The web interfaces associated with compensation collection and insights serve traffic over HTTPS. We also encrypt the communication between LinkedIn's web servers and datacenters (through intermediate layers) using TLS. We also secure the data storage layers using a combination of encryption (explained above) and authentication (e.g., only allow connections to the data store from trusted services running on (preconfigured) trusted hosts).\\

\subsubsection{Preventing Timestamp Based Inference Attacks}\label{sec:timestampjoin}
We next describe the mechanisms supported by our system for preventing timestamp based inference attacks. These attacks rely on the ability to join the de-identified compensation data with data containing member identity, on the timestamp information. For example, page view and other event logs could contain information on when a member accessed the page associated with the compensation collection web interface. Hence, it is not desirable for the exact timestamp to be retained with a submitted compensation entry in the de-identified data. One possible mechanism is to perturb the timestamp by adding a random time period drawn from a certain distribution. For example, we could add a random delay up to say, 48 hours to the true timestamp, and make the data available for processing only after this delay. While simple and elegant to implement, this approach may not provide sufficient protection when the submissions arrive infrequently in a cohort, for example, once every week. Hence, our design also supports another mechanism based on $k$-Anonymity wherein the modification to the timestamp is a function of other submissions in a cohort. The key idea is to define a hierarchy of timestamps (e.g., second $\rightarrow$ minute $\rightarrow$ hour $\rightarrow$ date $\rightarrow$ week $\rightarrow \ldots$), and generalize the timestamp for each compensation entry in a cohort to the granularity at which there are at least $(k-1)$ other entries with the same timestamp value. However, since our approach needs to be incremental in practice, we can achieve this property by processing entries within a cohort in batches of size $k$, generalizing to a common timestamp within each batch, and making additional data available for offline processing only in such incremental batches. Processing in batches could also be used to achieve incremental privacy protection, since the insight for a cohort would not be sensitive to a single new submission but only batches of $k$ or more submissions.

\subsubsection{Protecting Against Over-representation}\label{sec:submissionlimits}
The quality of the compensation insights depends crucially on ensuring that the submissions are representative of the underlying cohort. Although we reach out to randomly sampled members from each cohort as discussed in \S\ref{sec:targeting}, we need to also ensure that a member cannot submit the compensation data multiple times, or too often to avoid over-representation of certain members. 
For this reason and also for enabling the give-to-get model, we maintain a data store containing when a member submitted compensation information, and query this store to constrain how often a member can re-submit the compensation data. We can achieve this goal in different ways: (1) by limiting the frequency (e.g., at most once every 12 months), (2) by limiting the frequency, customized for different industries/functions by taking into account how often pay changes, how often people change jobs, etc, or (3) based on a change in a member's profile (e.g., whenever the member adds a new job position, or updates the job description).
 \section{Experiments}\label{sec:exp}
We next present an experimental study of the LinkedIn Salary system, focusing on the the tradeoffs between privacy and modeling requirements.

\subsection{Experimental Setup}\label{sec:expsetup}
As stated earlier, our system has been deployed in production at LinkedIn, initially for collection of compensation data from LinkedIn members and later for powering the publicly launched LinkedIn Salary product. We study the distribution of responses over time and investigate the tradeoffs between privacy guarantees and data available for computing compensation insights. Our experiments are performed using more than one year of compensation submission history data collected from over 1.5 million LinkedIn members, across three countries (USA, Canada, and UK). Note that we use just the submission history data, and not the compensation data itself for these experiments. Since LinkedIn Salary uses a give-to-get model, we need to keep track of the submission history for each member (as discussed in \S\ref{sec:collection}). An extensive study and evaluation of the compensation insights and the associated statistical modeling components is presented in~\cite{KAZA17}.

\begin{figure}
\centering
\includegraphics[width=\columnwidth]{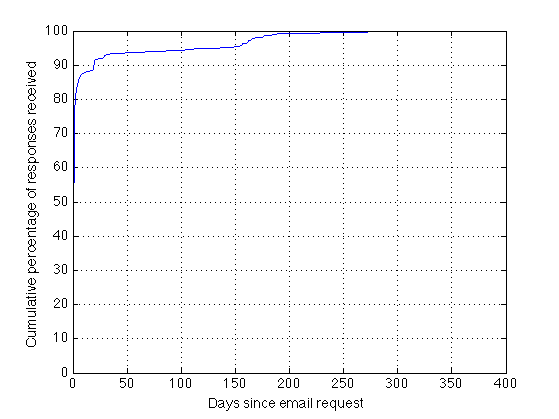}
\caption{The distribution of compensation submissions as a function of time lapsed since the email request}
\label{fig:respovertime}
\end{figure}

\subsection{Studying the Distribution of Compensation Submissions}\label{sec:distributionexp}
We first study how soon members respond upon receiving the emails requesting them to provide the compensation information. Figure~\ref{fig:respovertime} shows the cumulative percentage of responses received as a function of the time lapsed. In this figure, the X-axis denotes the number of days that have lapsed since the email request. For each possible value $d$ of the number of days lapsed, we compute the corresponding number of member submission responses, and thereby obtain the cumulative percentage of responses that were received within $d$ days of email request, plotted in Y-axis. We observe that 80\% of responses were received within one day of email request, suggesting that batching together responses to achieve greater privacy is not likely to lead to significant delay or withholding of data for modeling purposes. Further, more than 85\% of all responses were received within first 5 days of email request. The responses took considerable time in certain cases: about 5\% of responses were received after 135 days! 

\subsection{Studying the Tradeoffs between Privacy and Modeling Needs}\label{sec:tradeoffsexp}
\begin{figure}
\centering
\includegraphics[width=\columnwidth]{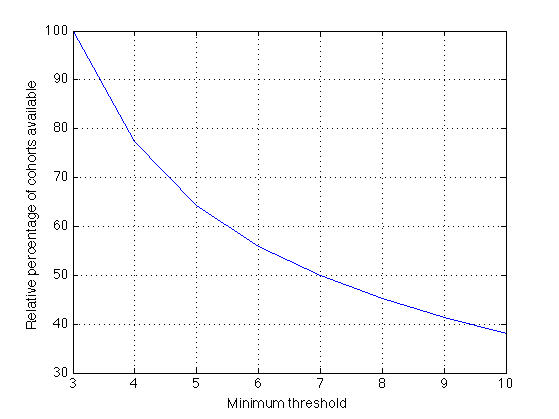}
\caption{Relative percent of cohorts that are available vs. the minimum threshold}
\label{fig:perc_cohorts}
\end{figure}

As discussed in \S\ref{sec:grouping}, a cohort needs to contain a minimum number $k$ of entries before the cohort and the associated compensation data is made available for offline data processing. By varying $k$, we obtain a tradeoff between the level of privacy protection and the amount of data available for modeling. Figure~\ref{fig:perc_cohorts} presents the relative percent of title-country-region cohorts that are available as a function of the minimum threshold, $k$. Note that we only used cohorts with at least 3 entries for this analysis. We observe that only about 65\% of the cohorts will be available with a threshold of 5 and 38\% of the cohorts with a threshold of 10, compared to a threshold of 3. In other words, out of the cohorts with at least 3 entries, about one-third have either 3 or 4 entries each; slightly less than one-third have between 5 and 9 entries each; and the rest have 10 or more entries each.

\begin{figure}
\centering
\includegraphics[width=\columnwidth]{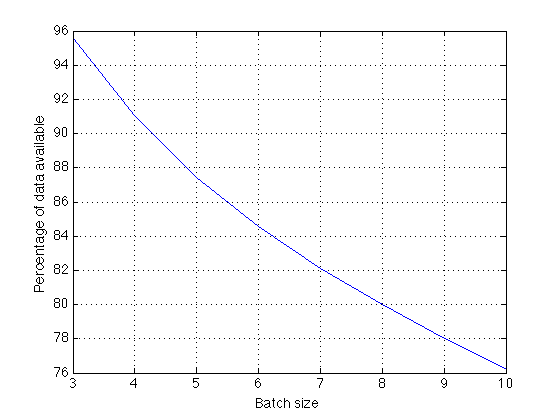}
\caption{Percent of data available vs. the batch size}
\label{fig:perc_data}
\end{figure}

Next, we studied the effect of processing entries within a cohort in batches of size $k$ (in addition to requiring a minimum threshold of $k$), and generalizing to a common timestamp within each batch. As discussed in \S\ref{sec:timestampjoin}, processing in batches helps to protect against timestamp based inference attacks and to also achieve incremental privacy protection. However, this results in some data being delayed or unavailable for offline processing. For example, if there are currently 8 submissions in a cohort and the batch size is 5, then the 6th, 7th, and 8th entries would not be made available. Figure~\ref{fig:perc_data} shows the percent of data that is available as a function of the batch size, $k$. For each $k$, we compute the number of entries available due to processing in batches (aggregated across cohorts) divided by the total number of entries (aggregated across cohorts).
Again, we only considered title-country-region cohorts that have at least 3 submissions. We see that about 4.5\% of the submissions would be withheld with a batch size of $k = 3$ (compared to none in the absence of batching, since we only consider cohorts with at least 3 entries), about 13\% of the submissions would be withheld with a batch size of $k = 5$, and about 24\% of the submissions would be withheld with a batch size of $k = 10$.

\begin{figure}
\centering
\includegraphics[width=\columnwidth]{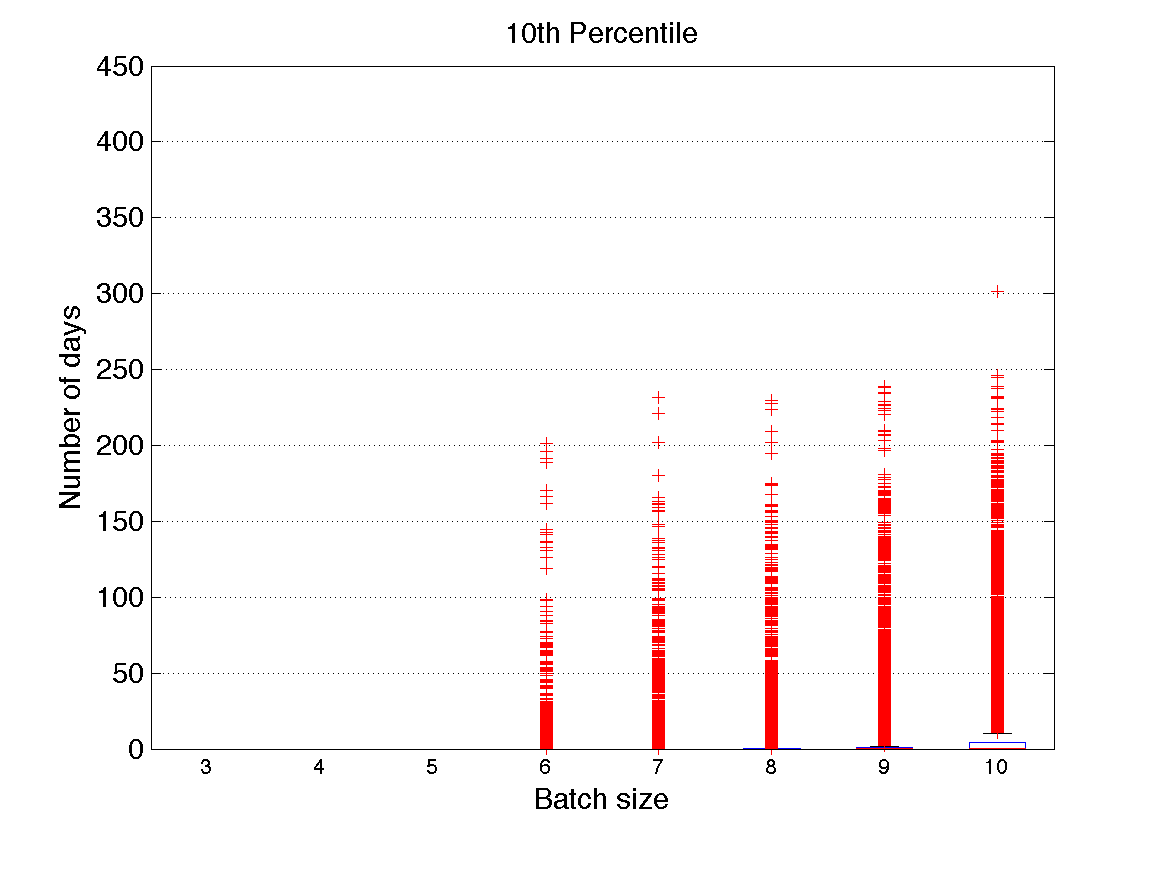}
\caption{10th percentile of the delays introduced by batching}
\label{fig:10p}
\end{figure}

\begin{figure}
\centering
\includegraphics[width=\columnwidth]{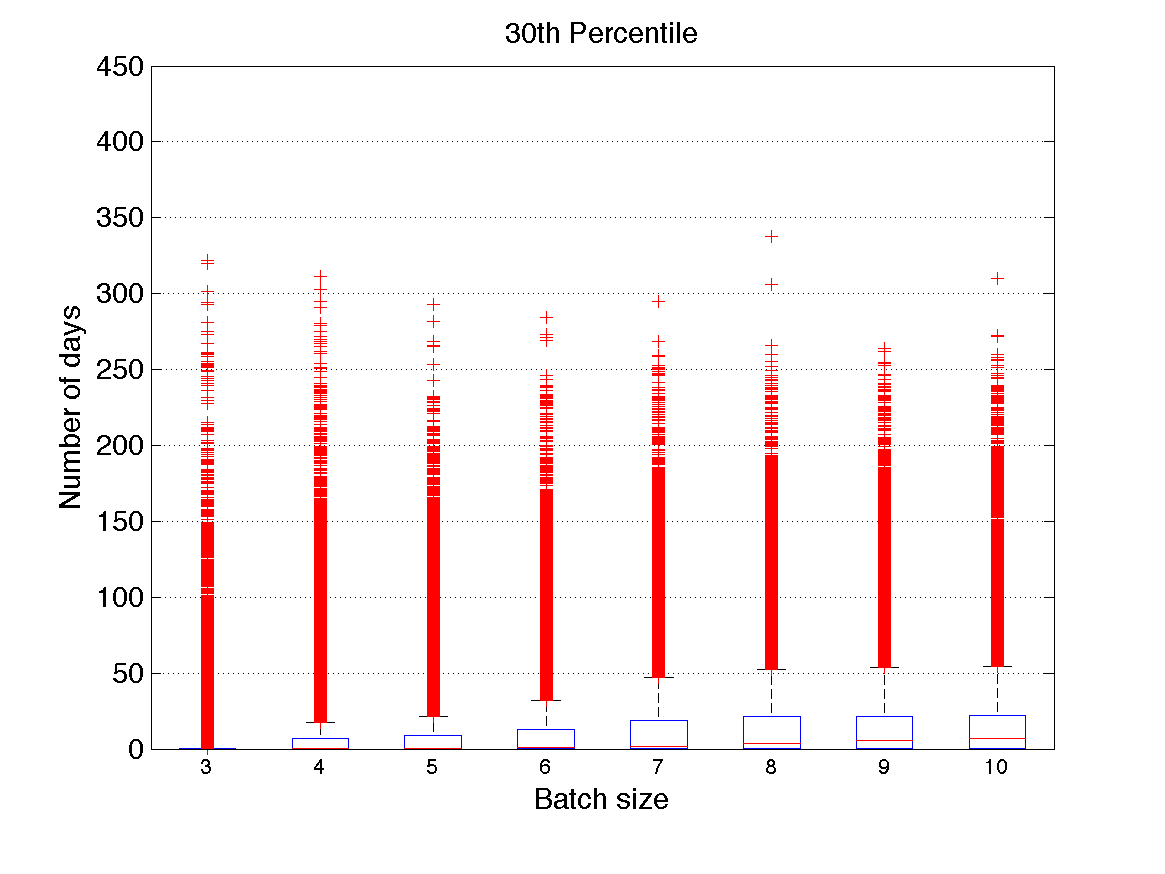}
\caption{30th percentile of the delays introduced by batching}
\label{fig:30p}
\end{figure}

\begin{figure}
\centering
\includegraphics[width=\columnwidth]{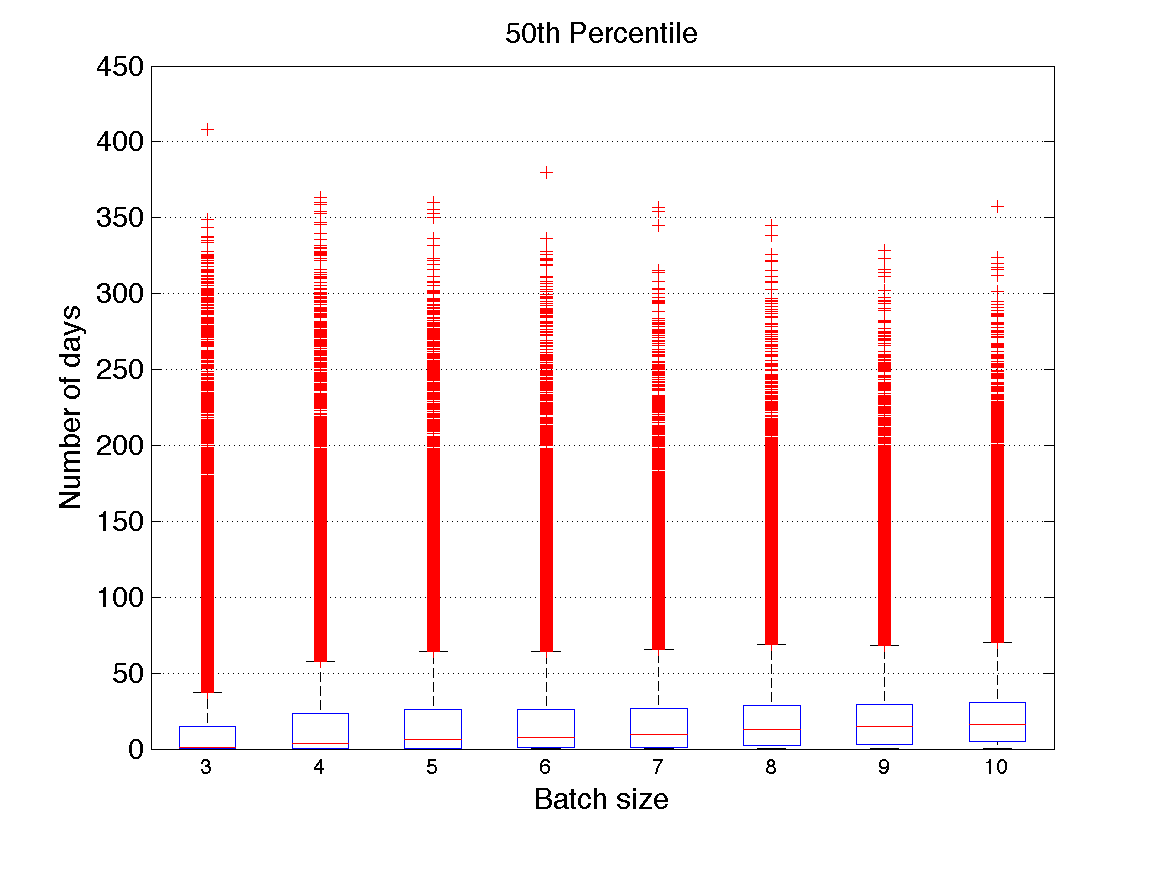}
\caption{50th percentile of the delays introduced by batching}
\label{fig:50p}
\end{figure}

\begin{figure}
\centering
\includegraphics[width=\columnwidth]{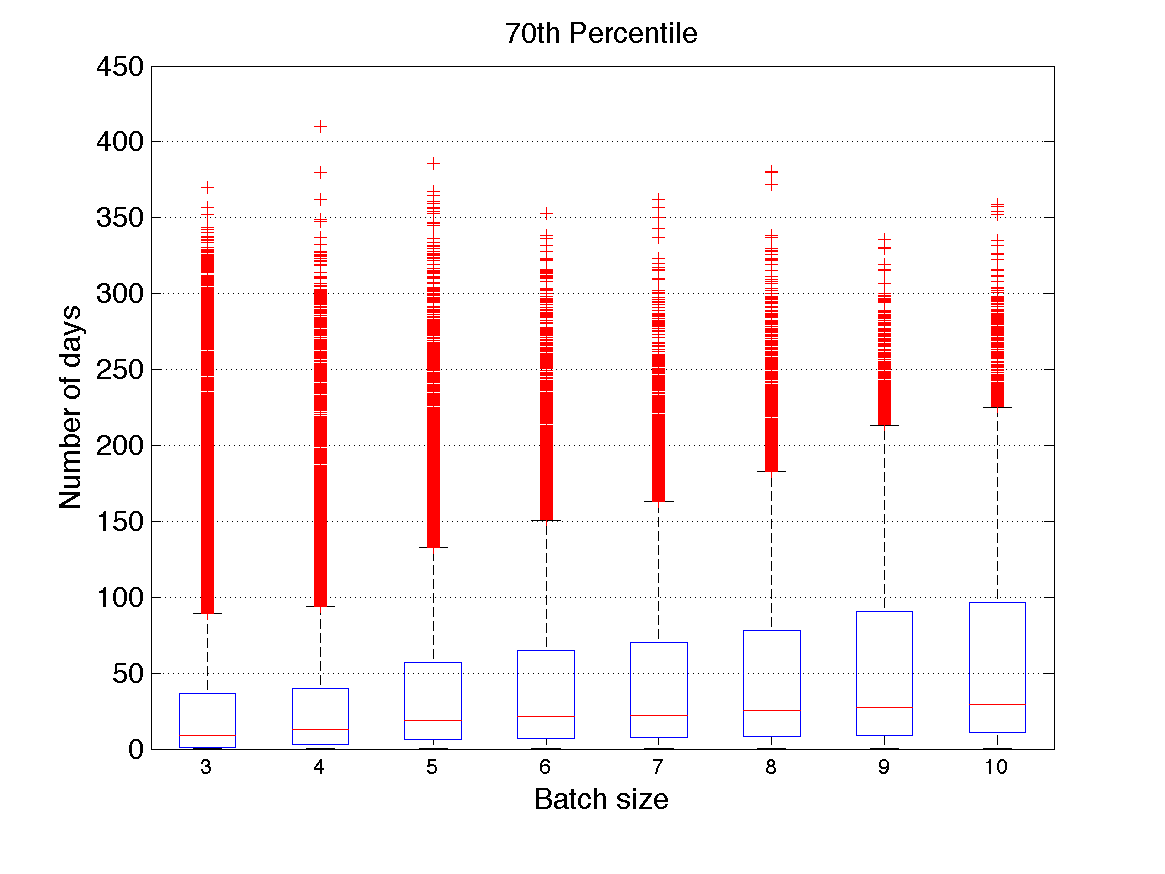}
\caption{70th percentile of the delays introduced by batching}
\label{fig:70p}
\end{figure}

\begin{figure}
\centering
\includegraphics[width=\columnwidth]{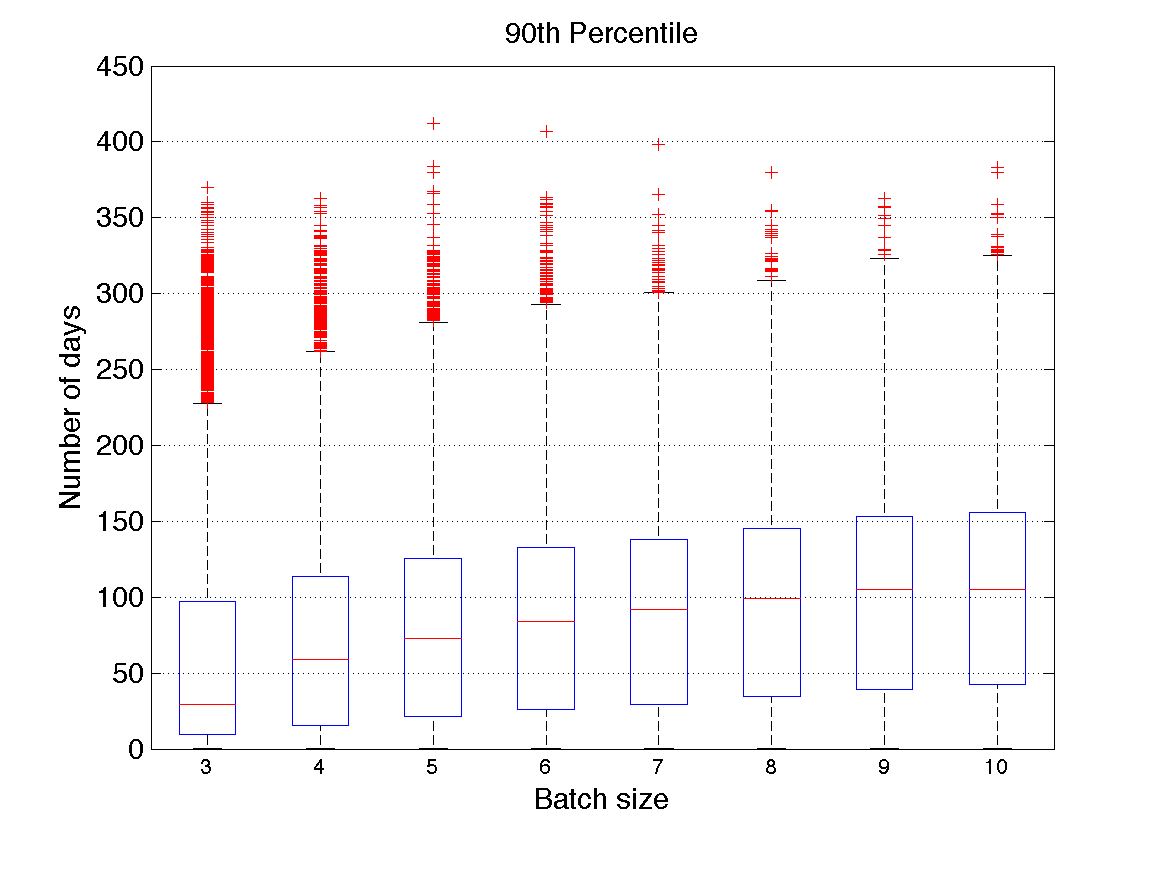}
\caption{90th percentile of the delays introduced by batching}
\label{fig:90p}
\end{figure}

We also investigated the actual delays that could emerge as a result of processing data in batches. Our results are presented in Figures~\ref{fig:10p}--\ref{fig:90p}. We calculated these delays as follows. For each title-country-region cohort, we first compute the delays introduced for certain submissions as a result of processing in batches, limiting to those submissions that would get processed. For example, if a cohort has 8 entries submitted on days $1, 2, \ldots, 8$ respectively, and the batch size is 5, then we would ignore the last three entries, and associate a delay of 4 days, 3 days, 2 days, 1 day, and 0 day respectively for the first five entries. We then sort the delay values within each cohort, and compute different percentiles (for instance, the median delay (50th percentile) would be 2 days in the above example). For each batch size and each percentage, we aggregate the corresponding delay percentile values across all cohorts, and show the distribution using a box-and-whisker plot. The median across all cohorts is shown as a red horizontal line, with the box boundaries corresponding to $q_1$ (25th percentile) and $q_3$ (75th percentile) respectively. The ends of the whiskers correspond to the lower limit, computed as max$(0, q_1 - 1.5(q_3 - q_1))$ and the upper limit, computed as $q_3 + 1.5(q_3 - q_1)$ respectively. The points outside this range are plotted as outliers (each with a red plus sign). For example, we can infer from Figure~\ref{fig:30p} that, for a batch size of 8, the 30th percentile delay values range from 0 to 24 days for three-fourth of the cohorts, with a median of 4 days. Similarly, Figure~\ref{fig:50p} shows that the median delay for three-fourth of the cohorts is less than about five weeks for batch sizes up to 10, although the median delay could be as high as 300 to 400 days for a few outlier cohorts that receive relatively fewer and infrequent submissions.

 \section{Lessons Learned in Practice}\label{sec:lessons}
Our architecture and design choices evolved over the course of the project, by taking into account various practical considerations. The product design and user interface has been evolving over time based on feedback from end users as well as focus group studies, and consequently, our architecture had to be designed to enable rapid iterations, for example, through easy creation and modification of cohort types, and flexible design of key-value stores. From time to time, we also performed ``bootstrapping'' to regenerate the historical output data due to changes in some system components, e.g., due to software bugs in the system implementation, changes to the various services that are invoked by the system, and changes to the product design such as addition of new types of insights. Our privacy model choices were also influenced by the unique product needs as well as limitations of our setting, as discussed in \S\ref{sec:related}. As discussed in \S\ref{sec:grouping}, we first considered storing and processing the encrypted (but sensitive) data in LinkedIn's Hadoop ecosystem, but chose the de-identification approach instead due to the security limitations of Hadoop and HDFS. During deployment of our system, we realized that it is not only necessary to have separate services for the encryption of member attributes and the compensation data, but also that these services need to be deployed on different machines. Otherwise, an attacker could get access to a single machine containing both services, thereby being able to decrypt both the member attribute data and the compensation data, which defeats our goals. We used custom deployment tools to ensure that these services are deployed in different machines.

 \section{Related Work}\label{sec:related}

{\em Salary Information Products}: There are several commercial services offering information pertaining to compensation and benefits. For example, Glassdoor~\cite{GlaPr1} offers a comparable service, while PayScale~\cite{PayOne} collects individual salary submissions, offers free reports for detailed matches, and sells compensation information to companies.  The US Bureau of Labor Statistics~\cite{BlsOvr} publishes a variety of statistics on pay and benefits.

{\em Privacy}: Preserving user privacy is important when collecting compensation data, considering its sensitive nature. There is rich literature in the field of privacy-preserving data mining spanning different research communities (e.g.,~\cite{samarati2001protecting, Swe02, adam1989security, agrawal2000privacy, evfimievski2003limiting, kantarcioglu2004privacy, vaidya2002privacy, kenthapadi2005simulatable, aggarwal2005two, machanavajjhala2007diversity, li2007t}), as well as on the limitations of simple anonymization techniques (e.g.,~\cite{backstrom2007wherefore, narayanan2008robust, su2017anonymizing}). Based on the lessons learned from the privacy literature, we first attempted to make use of rigorous privacy techniques such as differential privacy~\cite{DKM+06,DMNS06} in our problem setting. However, we soon realized that these are not applicable in our context for the following reasons: (1) the amount of noise to be added to the quantiles, histograms, and other insights would be very large (thereby depriving the compensation insights of their reliability and usefulness), since the worst case sensitivity of these functions to any one user's compensation data could be large, and (2) the insights need to be provided on a continual basis with the arrival of new data points. Although there is theoretical work on applying differential privacy under continual observations~\cite{CSS11,DNPR10}, we have not come across any practical implementations or applications of these techniques. We also explored approaches similar to recent work at Google~\cite{EPK14} and Apple~\cite{Gre16} on privacy-preserving data collection at scale that focuses on applications such as learning statistics about how unwanted software is hijacking users' settings in Chrome browser and discovering the usage patterns of a large number of iOS users for improving the touch keyboard respectively. These approaches are (or seem to be) built on the concept of randomized response~\cite{War65} and require response from typically hundreds of thousands of users for the results to be useful. In contrast, even the larger of our cohorts contain only a few thousand data points, and hence these approaches are not applicable in our setting.

{\em Survey Techniques}: There is extensive work on traditional statistical survey techniques~\cite{groves2011survey,jessen1978statistical}, as well on newer areas such as web survey methodology~\cite{callegaro2015web}. See~\cite{Ber05} for a survey of non-response bias challenges, and~\cite{bethlehem2010selection} for an overview of selection bias. 

 \section{Conclusions and Future Work}\label{sec:conclusion}
We presented the design and architecture of LinkedIn Salary, a system for securely collecting compensation information from LinkedIn members and providing structured compensation insights to job seekers. We highlighted unique challenges associated with privacy and security, and described how we addressed them using techniques such as encryption, access control, de-identification, aggregation, and thresholding. We demonstrated the tradeoffs between privacy and modeling needs through an experimental study with more than one year of compensation submission history data collected from over 1.5 million LinkedIn members. We also discussed the design decisions and tradeoffs while building our system, and the lessons learned from the production deployment of this system at LinkedIn.

An interesting direction to extend this work is to investigate approaches for improving the balance between privacy and modeling needs. For instance, we could explore the applicability of provably privacy-preserving machine learning approaches (e.g.,~\cite{chaudhuri2011differentially,papernot2016semi}) in our setting. Such explorations would require a redesign of the system, wherein we perform modeling on a secure cluster of servers with access to production databases and build richer prediction models that make use of more discriminating features beyond those available post de-identification. We would also like to incorporate outlier detection during submission stage by using user profile and behavioral features. Further research directions for improving the modeling components and for studying compensation data towards improving the efficiency of career marketplace are discussed in~\cite{KAZA17}. More broadly, by presenting the practical challenges encountered and the lessons learned during the design and deployment of privacy mechanisms for an emerging internet application, we hope to stimulate new research directions in privacy-aware computing.
 
\section*{Acknowledgments}
The authors would like to thank all other members of LinkedIn Salary team for their collaboration for deploying our system as part of the launched product, and
Stephanie Chou,
Tim Converse,
Tushar Dalvi,
Anthony Duerr,
David Freeman,
Joseph Florencio,
Ashish Gupta,
David Hardtke,
Parul Jain,
Prateek Janardhan,
Santosh Kumar Kancha,
Rong Rong,
Ryan Sandler,
Cory Scott,
Ganesh Venkataraman,
Liang Zhang,
and
Lu Zheng
for insightful feedback and discussions.

{
\bibliographystyle{abbrv}
\bibliography{paper}
}
\end{document}